\documentclass[twocolumn]{aastex631}
\usepackage{textcomp, gensymb}

\shorttitle{The Hottest Neptunes Orbit Metal-Rich Stars}
\shortauthors{Vissapragada \& Behmard}
\graphicspath{{./}{}}

\begin{document}

\title{The Hottest Neptunes Orbit Metal-rich Stars}

\email{svissapragada@carnegiescience.edu}
\email{abehmard@flatironinstitute.org}

\author[0000-0003-2527-1475]{Shreyas~Vissapragada}
\altaffiliation{Both authors contributed equally to this work.}
\affiliation{Carnegie Science Observatories, 813 Santa Barbara Street, Pasadena, CA 91101, USA}

\author[0000-0003-0012-9093]{Aida~Behmard}
\altaffiliation{Both authors contributed equally to this work.}
\affiliation{Center for Computational Astrophysics, Flatiron Institute, 162 Fifth Avenue, New York, NY 10010, USA}
\affiliation{American Museum of Natural History, 200 Central Park West, Manhattan, NY 10024, USA}

\begin{abstract}
The Neptune desert is no longer empty. A handful of close-in planets with masses between those of Neptune and Saturn have now been discovered, and their puzzling properties have inspired a number of interesting theories on the formation and evolution of desert-dwellers. While some studies suggest that Neptune desert planets form and evolve similarly to longer-period Neptunes, others argue that they are products of rare collisions between smaller planets, or that they are the exposed interiors of giant planets (i.e., ``hot Jupiters gone wrong''). These origin stories make different predictions for the metallicities of Neptune desert host stars. In this paper, we use the homogeneous catalog of stellar metallicities from Gaia Data Release 3 to investigate the origins of Neptune desert dwellers. We find that planets in the Neptune desert orbit stars that are significantly more metal rich than the hosts of longer-period Neptunes ($p = 0.0016$) and smaller planets ($p = 0.00014$). In contrast, Neptune desert host star metallicities are statistically indistinguishable from those of hot Jupiter host stars ($p = 0.55$). Therefore, we find it relatively unlikely that Neptune desert planets formed and evolved similarly to longer-period Neptunes, or that they resulted from collisions between smaller planets, at least without another metallicity-selective process involved. A more straightforward explanation for this result is that planets in the desert truly are the exposed interiors of larger planets. Atmospheric spectroscopy of Neptune desert worlds may therefore provide a rare glimpse into the interiors of giant exoplanets.
\end{abstract}

\section{Introduction} \label{sec:intro}
The ``Neptune desert'' (or ``sub-Jovian desert'') is a deficit of extrasolar planets between the masses of Neptune and Saturn with orbital periods shorter than a few days \citep{Szabo2011, Lundkvist2016, Mazeh2016}. The paucity of short-period planets at intermediate sizes is an important probe of planetary evolution. The formation of this desert may be partially due to rapid atmospheric loss: at such close distances to their host stars, Neptune-mass planets experience strong photoevaporation driven by high X-ray and extreme ultraviolet (XUV) irradiations, and envelope loss can also be greatly accelerated by Roche-lobe overflow \citep{Ionov2018, Owen2018, Koskinen2022, Thorngren2023}. Near the upper edge of the desert, planets are probably too massive for substantial photoevaporation, so the dearth of sub-Jovian planets in this region is more likely a relic of planetary formation and/or migration \citep{Vissapragada2022, Lazovik2023}. In particular, high-eccentricity migration results in a period-dependent boundary that reflects the minimum periastron distance during migration \citep[below which planets would be tidally disrupted;][]{Matsakos2016, Owen2018}. 

Over the past five years, a number of planets have been confirmed deep within the Neptune desert. These planets challenge conventional theories of planet formation and evolution. Remarkably, some of these ``ultrahot Neptunes'' are relatively low density ($\lesssim3$~g~cm$^{-3}$), with voluminous envelopes of hydrogen and helium \citep[e.g., LTT 9779 b and TOI-3261 b;][]{Jenkins2020, Nabbie2024}. These planets have somehow resisted total envelope loss despite their relatively low core masses (here, ``core'' refers to the bulk non-gaseous part of the planet). In stark contrast, some ultrahot Neptunes also have incredibly high densities (reaching 10~g~cm$^{-3}$ in some cases), with small envelopes and core masses greater than 30 $M_\Earth$, making them quite resilient to atmospheric photoevaporation \citep[e.g., TOI-849 b, TOI-332 b, and TOI-1853 b;][]{Armstrong2020, Naponiello2023, Osborn2023}. Because planetary cores are thought to undergo runaway gas accretion at much smaller masses \citep[roughly 10 $M_\Earth$, e.g.][]{Pollack1996, Lee2019}, it is surprising to observe gas-poor planets with such high core masses. 
\begin{figure*}
    \centering
    \includegraphics[width=0.85\textwidth]{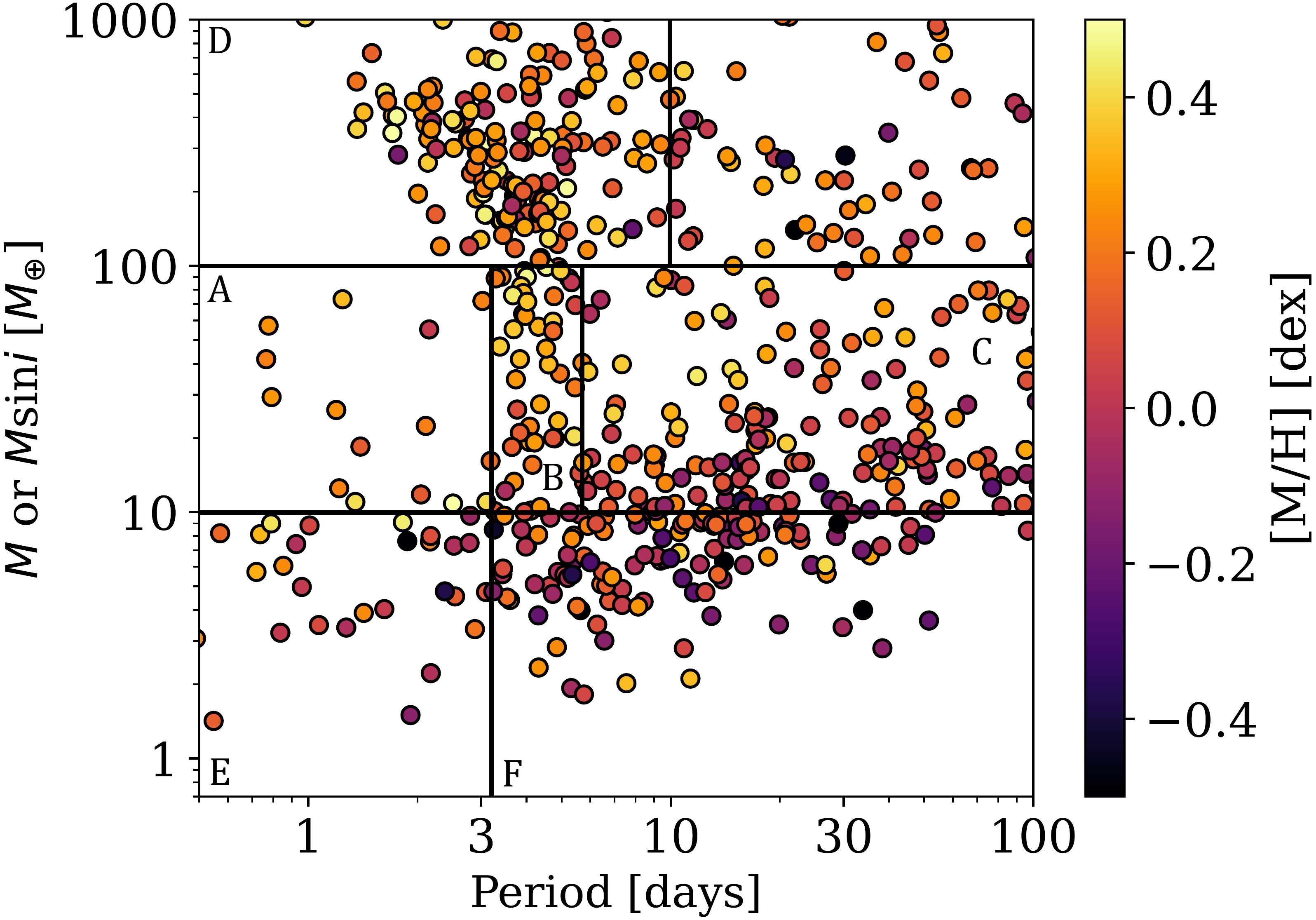} 
\caption{The mass-radius diagram of all confirmed planets from the NASA Exoplanet Archive with reported mass measurements and \textsf{GSP-Spec} metallicities from Gaia DR3. All planets are colored by the host star \textsf{GSP-Spec} metallicities. Different planet populations of interest are denoted by letters. A denotes the Neptune desert (average [M/H] = 0.25 dex), B the Neptune ridge (average [M/H] = 0.25 dex), C the Neptune savanna (average [M/H] = 0.10 dex), D the hot Jupiters (average [M/H] = 0.22 dex), E the small hot planets (average [M/H] = 0.07 dex), and F all small planets considered (average [M/H] = 0.03 dex).}
\label{desertfig}
\end{figure*}

A number of theories have been proposed for the origins of these unique Neptune desert planets, which we organize here into three categories depending on initial size/mass:
\begin{enumerate}
    \item ``Same size'': it is possible that Neptune desert planets formed and evolved similarly to longer-period Neptunes. As suggested by e.g., \citet{Armstrong2020} and \citet{Osborn2023}, core-dominated planets in the desert may have evaded runaway gas accretion by opening gaps in the natal gas disks \citep{Crida2006, Duffell2013}, or by forming after the gas disk dissipated \citep{Lee2019}. Planets that retained substantial envelopes (e.g., LTT 9779 b) may have escaped substantial atmospheric mass loss due to unusually low XUV irradiations and/or high atmospheric metallicities \citep{FernandezFernandez2024, Radica2024, Vissapragada2024}.
    \item ``Bottom up'': planet-planet collisions may produce planets in the Neptune desert, particularly those with high densities and core masses. In this scenario, two or more planets collide and merge to produce a single planet with an atypically large core mass \citep{Liu2015, Ginzburg2020, Ogihara2021, Naponiello2023}. While these events are rare \citep[see e.g.,][who simulated giant-impact outcomes and found that metal-rich giant-impact remnants were expected $<1\%$ of the time]{Cambioni2024}, desert dwellers are similarly rare relative to the small planet population \citep{Dai2021, Castro-Gonzalez2024}. However, impact luminosities may be large enough to unbind large gaseous envelopes, so this mechanism seems less favorable for explaining lower-density Neptune desert planets like LTT 9779 b. 
    
    \item ``Top down'': desert dwellers may have started their lives as giant planets that lost their envelopes, i.e., they may be ``hot Jupiters gone wrong.'' Photoevaporation alone is not efficient enough to remove most of the envelope of a gas giant progenitor, even at these close distances \citep{Vissapragada2022, Naponiello2023, Lazovik2023, Osborn2023}. If the progenitor was originally a short-period gas giant, Roche-lobe overflow could plausibly result in substantial envelope loss \citep{Jackson2016, Jenkins2020, Koskinen2022, Nabbie2024}. More catastrophic processes like envelope disruption during high-eccentricity migration could also remove the envelope of a progenitor giant planet \citep{Faber2005, Guillochon2011, Owen2018}.
\end{enumerate}

These theories make different predictions for the stellar metallicities of Neptune desert host stars. Because giant planets occur far more frequently around metal-rich stars \citep{Gonzalez1997, Fischer2005}, we expect that their host stars should mostly be metal rich if they formed top down and are the exposed interiors of giant planets. On the other hand, smaller planet occurrence rates are not as strongly correlated with host star metallicity \citep{Sousa2008, Ghezzi2010, Buchhave2012, Petigura2018}, so we are unlikely to observe a strong preference for metal-rich Neptune desert host stars if desert planets began at the same size or formed bottom up. Bottom-up formation may also require progenitor systems with multiple planets that can collide and merge. The host stars of mature multiplanet systems are observed to be either similar in metallicity \citep{MunozRomero2018, Weiss2018} or more metal poor \citep{Brewer2018, Anderson2021, RodriguezMartinez2023} to those of single-planet systems, so this requirement is unlikely to result in a preference for metal-rich Neptune desert hosts in the bottom-up formation scenario.

In this work, we show that the hottest Neptunes do indeed orbit stars that are more metal-rich than the Sun---similar to hot Jupiters, and different from other populations of smaller planets. In Section~\ref{sec:sample}, we describe our sample selection methodology along with the source of our stellar metallicities. In Section~\ref{sec:neptunes}, we show that the hottest Neptunes have almost exclusively been found around metal-rich stars, which would not be expected if these planets formed and evolved similarly to longer-period Neptunes or if they formed bottom-up. We show that our main results are robust to various methodological choices in Section~\ref{sec:robust}. Finally, we discuss the implications of this finding in Section~\ref{sec:conc}.

\section{Sample Selection} \label{sec:sample}

Our aim is to determine the metallicity distribution of Neptune desert planet host stars, and compare this distribution to those of longer-period Neptune hosts (testing the ``same-size'' theory), smaller planet hosts (testing the ``bottom-up'' theory), and larger planet hosts (testing the ``top-down'' theory). We defined our samples on the $P$$-$$M_p$ plane \citep[where the desert boundaries are sharpest; e.g.,][]{Mazeh2016} using the catalog of confirmed planets from the NASA Exoplanet Archive (for which false-positive rates are low). The implications of studying the desert in the mass-period plane rather than the radius-period plane are discussed further in Section~\ref{sec:robustrad}. We only considered planets with reported mass measurements and fractional $M_p$ or $M_p\sin i$ precisions better than 30\%. For planets with more than one reported mass measurement, we took the mass with the lowest associated error. 

To construct the Neptune desert sample, we selected planets with $P < 3.2$~d \citep[the period boundary determined by][]{Castro-Gonzalez2024}, and 10 $M_\Earth<M_p<$ 100 $M_\Earth$. The upper mass boundary distinguishes the Neptune desert from hot Jupiters \citep[e.g.,][]{Dawson2018}, and the lower mass boundary distinguishes the desert from the hottest rocky planets, for which formation pathways are likely quite different \citep{Aguichine2020, Dai2021, Johansen2022, Cambioni2024, Lin2024, Rubenzahl2024}. Our mass and period cuts are shown in Figure~\ref{desertfig} and select only planets deep within the Neptune desert, whereas earlier desert definitions from e.g., \citet{Mazeh2016} or \citet{Lundkvist2016} now include some planets at the desert edges, which may have formed and/or evolved differently \citep{Castro-Gonzalez2024}.

To test the ``same size'' theory, we compared the Neptune desert sample to planets with the same masses in the newly-detected Neptune savanna \citep{Bourrier2023} and ridge \citep{Castro-Gonzalez2024}. We used the orbital period boundaries from \citep{Castro-Gonzalez2024}: $3.2$ d $< P < 5.7$~d for the ridge and $5.7$~d~$<P<100$~d for the savanna. To test the ``bottom up'' theory, we compared the desert sample to the population of small planets ($M_p<$ 10 $M_\Earth$), considering subsamples of hot planets ($P < 3.2$~d) and a broader population of warm planets ($P < 100$~d). Finally, to test the ``top down'' theory, we compared the desert sample to hot Jupiters (100 $M_\Earth<M_p<$ 13.6 $M_\mathrm{J}$, $P < 10$~d). These comparison samples are all shown in Figure~\ref{desertfig}.

We then collected host star metallicities for each of these samples. The metallicities reported by the NASA Exoplanet Archive are from individual planet discovery papers, and are quite heterogeneous in provenance. We therefore decided to use a homogeneous metallicity catalog for our study. Like \citet{Yee2023}, we used the catalog of \textsf{GSP-Spec} metallicities reported by Gaia Data Release 3 \citep[DR3;][]{GaiaCollaboration2016, GaiaCollaboration2023}, which were homogeneously determined using data from the Gaia Radial Velocity Spectrometer \citep[$R\sim11,500$; ][]{Cropper2018}. We only used \textsf{GSP-Spec} metallicities with $<$0.1~dex uncertainties; the average metallicity uncertainty for stars in our sample is $\sim$0.02~dex. We also applied the color-dependent calibration determined by \citet{Yee2023} to debias the metallicities. Throughout this work we used the stellar mean metallicity [M/H] rather than the iron abundance [Fe/H], as [M/H] is both the base quantity reported by \textsf{GSP-Spec} \citep{Recio-Blanco2023}, and also the quantity calibrated by \citet{Yee2023}. All points in Figure~\ref{desertfig} are colored by the debiased Gaia metallicities. 

\section{The Hottest Neptunes Orbit Metal-rich Stars} \label{sec:neptunes}

Figure~\ref{desertfig} visually demonstrates that all planets in our Neptune desert sample orbit stars that are more metal-rich than the Sun.

\begin{figure*}[t]
    \centering
    \includegraphics[width=0.47\linewidth]{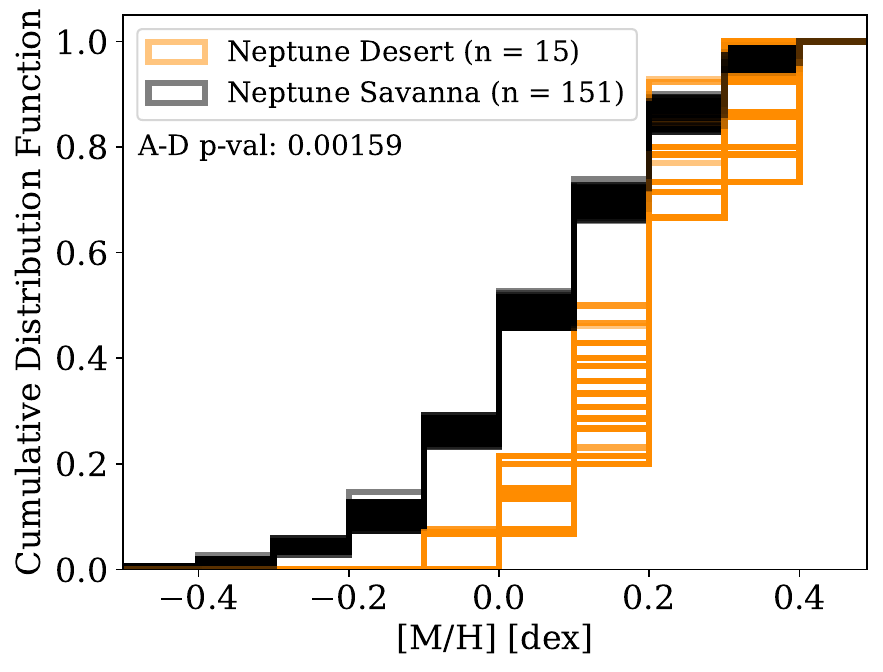}
    \includegraphics[width=0.47\linewidth]{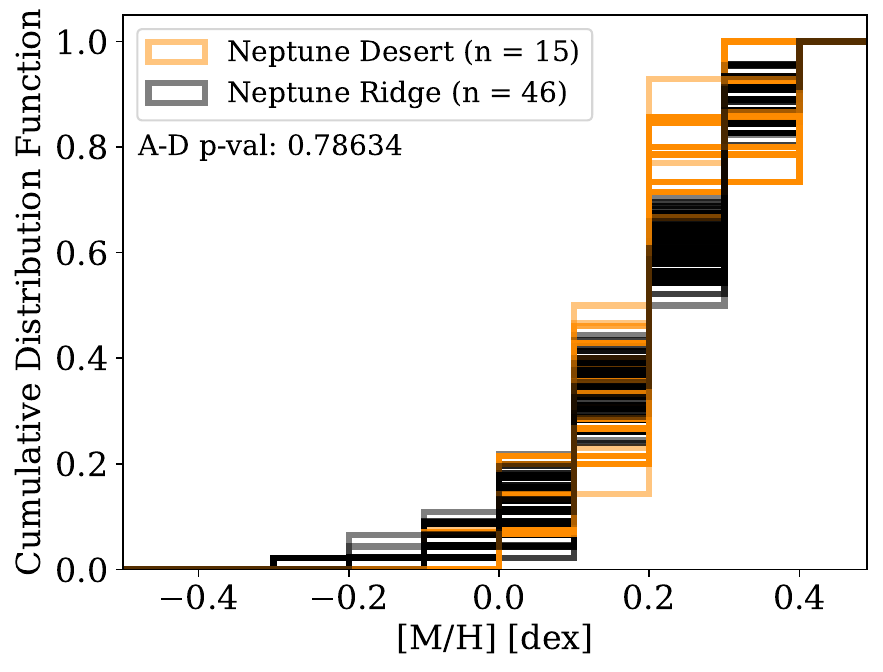}
    \includegraphics[width=0.47\linewidth]{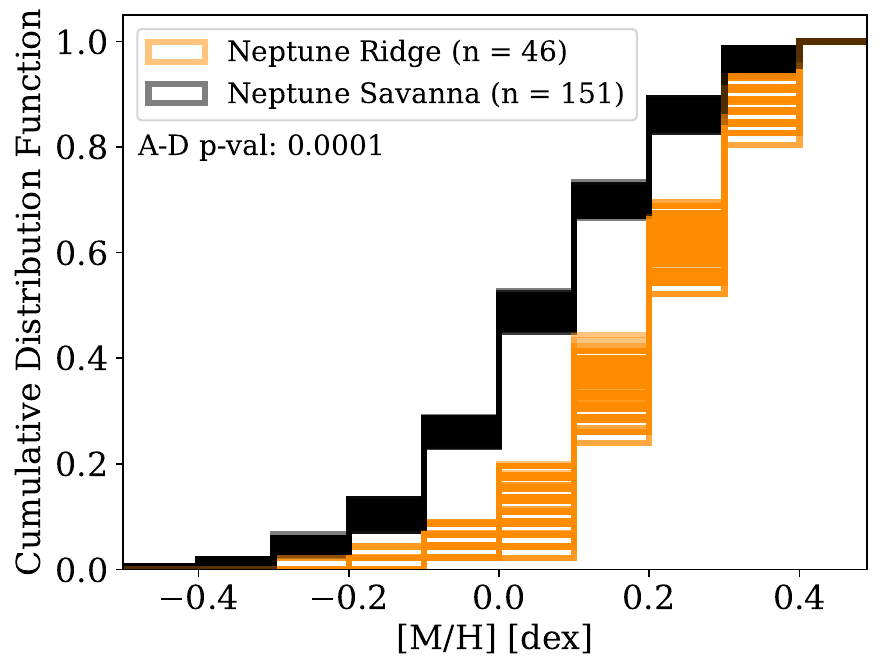}
    \includegraphics[width=0.47\linewidth]{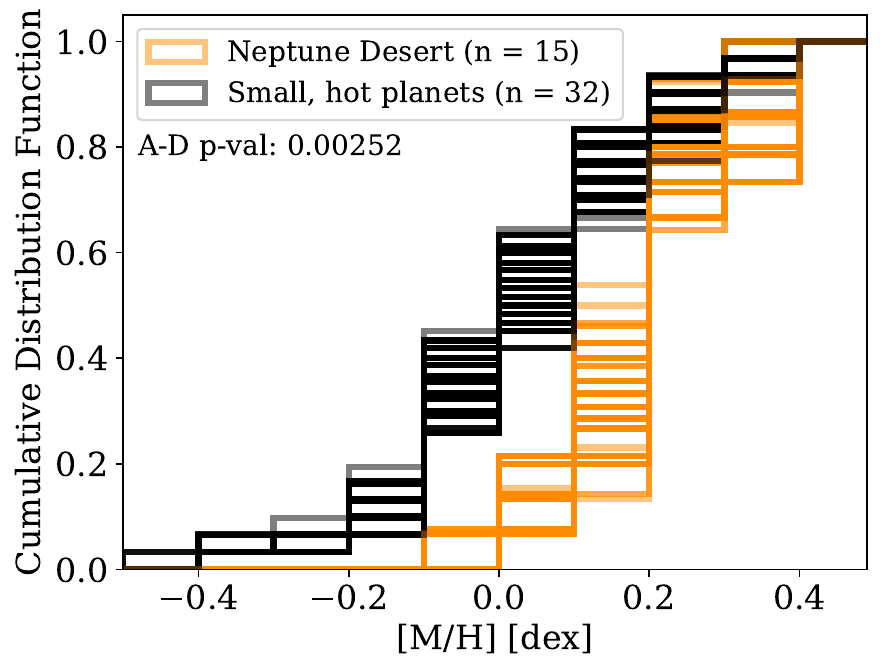}
    \includegraphics[width=0.47\linewidth]{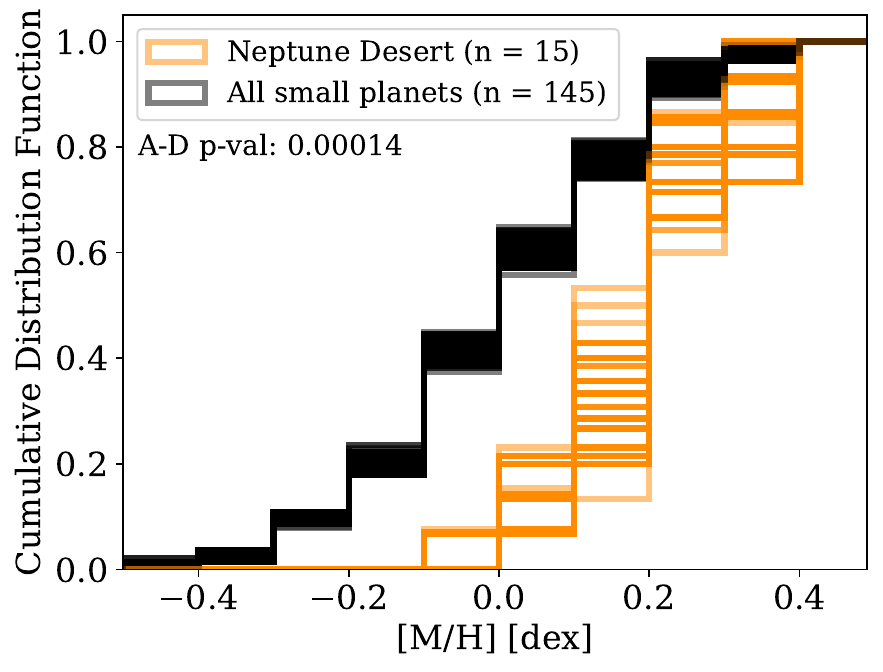}
    \includegraphics[width=0.47\linewidth]{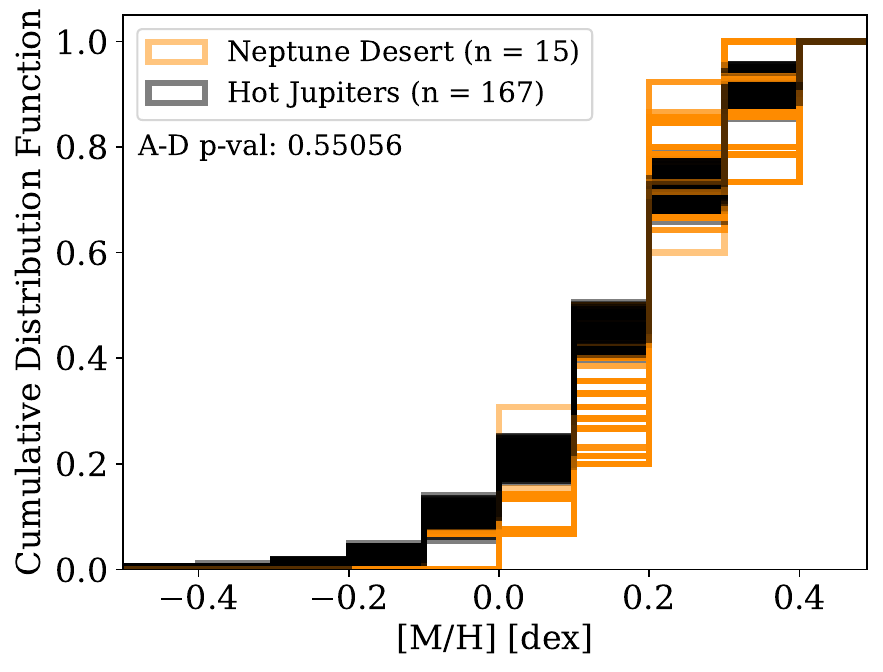}

\caption{Comparison of host star metallicity CDFs for different planet samples. From top left to bottom right, the panels correspond to: planets in the Neptune desert and savanna; desert and ridge (center); ridge and savanna; desert and small, hot planets; desert and all small planets; and desert and hot Jupiters. We resampled every stellar metallicity from a normal distribution (using the mean and standard deviation of the de-biased metallicity from \textsf{GSP-Spec}), and recompute the CDFs 1000 times to show the uncertainty.}
\label{neptunehist}
\end{figure*}

\subsection{Neptune Desert Hosts are More Metal-rich Than Neptune Savanna Hosts} \label{sec:main}

This would not be too surprising if longer-period Neptunes orbited similarly metal-rich stars. To test this possibility, in Figure~\ref{neptunehist} we compared the host star metallicities of the Neptune desert sample ($P<3.2$~d) with those of the Neptune savanna sample ($5.7$~d~$<P<100$~d). To calculate uncertainties in the metallicity CDFs, we resampled every stellar metallicity from a normal distribution (using the means and standard deviations of the debiased \textsf{GSP-Spec} metallicities) and recomputed the CDFs 1000 times  \citep[similar to bootstrap analyses in, e.g.,][]{RodriguezMartinez2023, Rosenthal2024}. The distributions appear quite different, even considering the uncertainties due to the relatively small sample size ($n = 15$) for the Neptune desert. To quantify the difference, we ran a two-sample Anderson-Darling (A-D) test on each of the 1000 pairs of resampled CDFs. We find a mean $p$ value of 0.0016, i.e., strong evidence that these two samples are not drawn from the same parent distribution of stellar host stars. Essentially, it is surprising that the Neptune desert host stars are so metal rich, given that there are quite a few metal-poor Neptune savanna hosts.

We applied the same analysis to compare the Neptune desert to the newly reported Neptune ridge \citep{Castro-Gonzalez2024}. This resulted in a mean $p$ value of 0.79, i.e., it is plausible that these subpopulations are sourced from the same parent population. Correspondingly, the CDFs in the middle panel of Figure~\ref{neptunehist} are visually indistinguishable. Finally, we repeated the analysis to compare the ridge hosts to the savanna hosts. We again find a low mean $p$ value of 0.0001, strong evidence for distinct host star populations between these two samples. 

These results demonstrate that the Neptune desert and ridge appear to be similar populations in terms of host star metallicity, but both appear quite distinct from the hosts of longer-period Neptune savanna planets. There is a growing body of literature suggesting that there are significant differences in dynamical architectures and bulk densities between short- and long-period Neptune systems \citep{Correia2020, Bourrier2023, Castro-Gonzalez2024b}. Our result is another important distinction between these two groups of planets.

\subsection{Neptune Desert Hosts are More Metal-rich than Small Planet Hosts}

Next, we assessed the likelihood that the hottest Neptunes formed ``bottom up'' by comparing the metallicities of Neptune desert hosts to those of smaller planets. If planets in the Neptune desert are the results of planet-planet collisions in systems of smaller planets, we would expect the metallicities of Neptune desert hosts to look similar to those of small planet hosts. Therefore, we compared the host star metallicity CDFs of the Neptune desert and small planet hosts with similar orbital periods ($M_p < 10$ $M_\Earth$, $P < 3.2$~d) using the same methodology as above. The result is shown in Figure~\ref{neptunehist}. These populations are distinct, with a mean $p$ value of 0.0025.

As a robustness test, we also compared the Neptune desert hosts to a broader population of small planets ($M_p < 10$ $M_\Earth$, $P < 100$~d) in Figure~\ref{neptunehist}. In this case, the mean $p$ value is even lower at 0.00014. It therefore seems unlikely that Neptune desert planets form from small planets collisions in multiplanet systems. If that were the case, we would expect to observe more metal-poor Neptune desert host stars than have actually been observed. Our results are in good agreement with previous population studies (mostly in the radius-period plane) that also found the metallicities of the hottest Neptune host stars to be distinct from those of smaller planet hosts \citep{Dong2018, Petigura2018, Dai2021}. 

\subsection{Neptune Desert Hosts have Similar Metallicities to Hot Jupiter Hosts}
Finally, we assessed the possibility that Neptune desert planets formed ``top down'' by comparing the metallicities of host stars with Neptune desert planets versus giant planets. If Neptune desert planets truly are the exposed interiors of gas giants, we would expect the metallicities of Neptune desert hosts to look similar to those of giant planet hosts. We therefore compared the Neptune desert sample to the sample of hot Jupiters ($100$ $M_\Earth < M_p < 13.6$ $M_\mathrm{J}$, $P < 10$~d). The host star metallicity CDFs for these populations are shown in Figure~\ref{neptunehist}. We find a mean $p$ value of 0.55, i.e., we cannot reject the hypothesis that these two distributions are drawn from the same parent population. This result is in good agreement with \citet{Dong2018}, who found that Neptune- and Jupiter-sized planets with 1~d $< P < 10$~d have similar host star metallicity distributions.

\section{Robustness Tests} \label{sec:robust}
\subsection{Radius-Period Space} \label{sec:robustrad}
Throughout this paper, we studied the desert in mass-period space using data from the NASA Exoplanet Archive, similarly to \citet{Mazeh2016}, \citet{Owen2018}, \citet{Szabo2019} and \citet{Szabo2023}. An advantage of using the Archive to define our sample is low false positive rates, but a disadvantage is the heterogeneity of planetary mass determinations across the community. For instance, if RV surveys were preferentially targeting metal-rich Neptune desert candidates, but not observing with the same metallicity preference in other populations, this would compromise the conclusions that we have drawn (although it seems unlikely that such a preference would be expressed across only a narrow range of mass-period space). Unfortunately, there is no single RV survey with a well-defined selection function across the full range of planetary masses and orbital periods studied here. Such a survey would certainly be a fruitful avenue for future work, and the high-resolution spectra obtained could also enable more precise stellar metallicity measurements than the debiased \textsf{GSP-Spec} values we used here.

Studying the desert in radius-period space allows for a larger and better-defined statistical sample from transit surveys, but there are a few challenges to this approach when it comes to studying host star metallicities in the desert. While the catalog of Kepler Objects of Interest (KOIs) is often a go to for such statistical analyses \citep[e.g.,][]{Mazeh2016, Castro-Gonzalez2024}, there are essentially no KOIs within the desert with well-measured metallicities, either from \textsf{GSP-Spec} or ground-based surveys, e.g., APOGEE, LAMOST, or CKS \citep{Dong2018, Petigura2018, Petigura2022}. On the other hand, TESS Objects of Interest (TOIs) are brighter and usually have \textsf{GSP-Spec} metallicities. There is virtually no difference in TESS light curve noise properties between metal-rich and metal-poor stars, so we can confidently rule out the metallicity detection biases that worry us when studying the desert in the mass-period plane \citep{Yee2023}. However, false positive rates are also quite high at the shortest orbital periods \citep{Sullivan2015}. Nevertheless, as a sanity check on our results, we repeated our analysis using the TOI catalog. As of 2024 October 25, there were 5437 unique TOIs listed with $P < 100$~d and dispositions of either ``Known Planet", ``Confirmed Planet", or ``Planet Candidate" according to the Exoplanet Follow-up Observing Program. We cross-matched the TOIs to Gaia using \textsf{astroquery}, and collected the \textsf{GSP-Spec} metallicities for all TOIs as above, resulting in 1292 objects.

Then, using the Neptune desert boundaries from \citet{Castro-Gonzalez2024}, we compared the metallicities of Neptune desert host stars and hosts of planets in the Neptune savanna, hot Jupiters, and small planets. We defined Neptune savanna planets as 4 $R_\Earth < R_p <$ 10 $R_\Earth$ and 5.7~d~$<P<100$~d, following the criteria for ``intermediate-sized'' planets from \citet{Castro-Gonzalez2024}. Hot Jupiters were defined as $R_p >$ 10 $R_\Earth$ and $P < 10$~d, and small planets as $R_p <$ 4 $R_\Earth$ and $P < 100$~d. Our results remain the same---the Neptune desert hosts are distinguishable from the Neptune savanna ($p = 0.03$) and small planet hosts ($p = 0.0003$), but not from the hot Jupiter hosts ($p = 0.77$). The $p$ values (especially for the desert-savanna comparison) are noticeably weaker as there are a few unconfirmed Neptune desert TOIs around relatively metal-poor stars. False positive rates are high in this orbital period regime, so these interesting candidates should be prioritized for follow-up observations.

\subsection{A Metal-poor Neptune Desert Host}
Some planet hosts do not have Gaia RVS spectra and were not included in our study due to a lack of \textsf{GSP-Spec} metallicities. The list of hosts without RVS spectra unfortunately includes NGTS-4, an apparently metal-poor star ([M/H]$=-0.28\pm0.10$~dex) which hosts a planet in the desert \citep{West2019}. To ensure that our results are robust to the omission of this interesting planet, we repeated our analysis from Section~\ref{sec:main} including the literature metallicity for NGTS-4. We found that our results are robust, although they weaken slightly ($p$ = 0.0063) for the desert to savanna comparison), as expected. As suggested by \citet{Dai2021}, a more precise metallicity constraint for NGTS-4 would be highly valuable, as it appears \textit{far} more metal poor than other Neptune desert hosts in our sample. 

\section{Conclusion} \label{sec:conc}
Three broad categories of origin stories have been proposed for planets in the Neptune desert: they formed similarly to longer-period Neptunes, they formed from smaller planets (through collisional growth), or they formed ``top down'' (through giant planet envelope loss). These three possibilities make different predictions for the metallicities of Neptune desert host stars, so we studied how they compare to the metallicities of other planet hosts with $P < 100$~d. We find the following:

\begin{enumerate}
    \item Planets in the Neptune desert are significantly ($p = 0.0016$) more metal-rich than longer-period planets in the Neptune savanna. It therefore seems unlikely that these populations formed and evolved similarly.
    \item In contrast, planets in the Neptune desert and the recently identified Neptune ridge (3.2 d $<$ $P$ $<$ 5.7 d) do not have distinguishable host star metallicity distributions ($p = 0.79$). It is plausible that these populations have similar origins.
    \item Neptune desert-dweller hosts are significantly more metal rich than small planet hosts, regardless of whether we consider small planets at similar orbital periods to the desert ($p = 0.003$) or out to 100~days ($p = 0.00014$). Therefore, it also seems unlikely that Neptune desert planets formed ``bottom up'', e.g., as the result of collisions between smaller planets.
    \item The metallicities of Neptune desert hosts and Hot Jupiter hosts are statistically indistinguishable ($p = 0.55$). It remains plausible that planets in the Neptune desert formed ``top down'', i.e., they are the exposed interiors of former gas giant planets. 
\end{enumerate}

Additionally, we found that these conclusions hold even if we use the full catalog of TOIs to define the Neptune desert, despite the relatively large false positive fraction at short orbital periods \citep{Sullivan2015}. Population studies of planet radii from Kepler arrived at similar conclusions for the ridge \citep[the Kepler desert is basically devoid of planets with well-measured host star metallicities;][]{Dong2018, Petigura2018, Petigura2022}. TESS has greatly increased the sample of desert dwellers with well-measured host star metallicities, allowing us to extend these previous studies into the desert. 

A straightforward explanation for our results is that the hottest Neptunes form ``top down'', i.e., these planets truly are the exposed interiors of gas giants. If hot Jupiters formed far from their host stars and arrived to their present positions via high-eccentricity tidal migration \citep{Dawson2018}, perhaps some experienced partial envelope disruption due to close pericenter passages \citep{Faber2005, Guillochon2011, Owen2018}. In this scenario, we may expect to see apparently lonely hot Neptunes accompanied by distant companions capable of driving dynamical migration \citep[e.g.,][]{Knutson2014, Ngo2016}. The majority of Neptune desert dwellers are observed to be isolated, seemingly in line with this idea. However, the occurrence of outer companions in these systems has not yet been systematically studied, and moreover this pathway seems unlikely for planets in the desert with nearby companions like TOI-4010b \citep{Kunimoto2023}, where high-eccentricity migration would have disrupted the delicate system architecture. Rapid envelope loss would need to have been dynamically quiescent in such cases, which could be achieved with Roche-lobe overflow of a progenitor hot Jupiter \citep{Valsecchi2014, Jackson2016, Jia2017, Nabbie2024}. In any case, if the hottest Neptunes did form top-down, then the low occurrence rate of Neptune desert planets reflects the relative rarity of giant planet envelope loss events. Future work focused on determining the frequency of Roche-lobe overflow/tidal disruption events capable of producing Neptune desert planets would therefore be highly valuable.

The concordance between the Neptune desert and hot Jupiter host star metallicity distributions is intriguing, but more complex evolution pathways are not strictly ruled out. Our finding that Neptune desert hosts are metal enriched could also be matched by same-size or bottom-up formation if the populations were modified by an additional metallicity-selective process that we have not considered here. Nevertheless, a natural explanation for our result is that the hottest Neptunes are indeed ``hot Jupiters gone wrong.'' If hot Jupiters have similar interior structures to Jupiter and Saturn \citep{Miguel2023, Helled2024}, we would expect the envelope metallicities of desert dwellers to be quite high in this scenario, reflecting the larger fraction of heavy elements expected in giant planet interiors. Transmission spectroscopy and/or phase curve observations can help experimentally verify this expectation \citep[e.g.,][]{Crossfield2020, Dragomir2020, Brande2022, Hoyer2023} and JWST is beginning to test this idea for a few standout Neptune desert systems \citep[including LTT 9779 b and TOI-849 b; ][GTO 1201, GO 3231, GO 5967]{Radica2024}. We encourage the community to continue pursuing these intriguing Neptune desert systems with JWST and other facilities. Such observations will offer rare insights into the interiors of giant planets if Neptune desert planets are indeed ``hot Jupiters gone wrong,'' as our results suggest.

\begin{acknowledgments}
We thank Sam Yee, Heather Knutson, Morgan Saidel, Mercedes L\'{o}pez-Morales, Jessica Spake, Michelle Kunimoto, Lucy Lu, Jiayin Dong, and James Owen for productive conversations. We also thank the anonymous referee for a helpful report. This research has made use of the NASA Exoplanet Archive, which is operated by the California Institute of Technology, under contract with the National Aeronautics and Space Administration under the Exoplanet Exploration Program. This work has made use of data from the European Space Agency (ESA) mission
Gaia (\url{https://www.cosmos.esa.int/gaia}), processed by the Gaia
Data Processing and Analysis Consortium (DPAC;
\url{https://www.cosmos.esa.int/web/gaia/dpac/consortium}). Funding for the DPAC
has been provided by national institutions, in particular the institutions
participating in the Gaia Multilateral Agreement.

\software{\texttt{numpy} \citep{numpy}, \texttt{matplotlib} \citep{matplotlib}, \texttt{pandas} \citep{pandas}, \texttt{scipy} \citep{scipy}, \texttt{scikit-learn} \citep{scikit-learn}, \texttt{astropy} \citep{astropy:2013, astropy:2018}}
\end{acknowledgments}

\bibliography{neptunes}{}
\bibliographystyle{aasjournal}

\end{document}